\documentclass[a4paper,11pt]{article}

\usepackage{amsmath, amsthm, amssymb}

\usepackage{fancyhdr}
\newcommand{\caH}{{\mathcal H}}

\newcommand{\caL}{{\mathcal L}}

\newcommand{\caS}{{\mathcal S}}

\newcommand{\bbC}{{\mathbb C}}

\newcommand{\bbN}{{\mathbb N}}

\newcommand{\bbR}{{\mathbb R}}

\newcommand{\bbZ}{{\mathbb Z}}

 \newtheorem*{thm}{Theorem}

\begin{document}

\title{On ground state phases of quantum spin systems}
\date{July 15, 2012}

\maketitle

\begin{center}
S. Bachmann \\
Department of Mathematics\\
University of California, Davis\\
Davis, CA 95616, USA \\
\textit{svenbac@math.ucdavis.edu}
\end{center}

\vspace{2cm}



Elucidating the ground state properties of a complex quantum system may be the most basic task in quantum theory, but it is in general a non-trivial one and one that provides a good insight in the system's behaviour, including aspects of the dynamics such as scattering theory. Among the prominent examples are the ground state properties of $N$-body quantum systems, the question of the existence of a ground state for quantum field theories and its construction, or the understanding of ground state spaces of quantum lattice systems. The renewed attention received by the latter in recent years arose in particular from the realization that non-trivial ground state spaces could be used as robust memories in quantum information.

Rather than studying the spectral problem of a given Hamiltonian in order to obtain very precise but quite specific properties of a model, a fruitful approach is to identify general classes of systems that share similar qualitative properties. In fact, the representation of a real system by a mathematical model follows that philosophy: the model is assumed to be in the same class as the physical system. Similarly, a large part of statistical physics over the past few decades has been devoted to the study of phase transitions, namely the transitions between qualitatively different thermal state(s) of a system -- the quintessential example being the melting of a solid to a liquid. Understanding qualitative transitions tuned by a parameter has found applications much further than for Gibbs states at different temperatures. Let me mention, e.g., Bose-Einstein condensation in the grand canonical ensemble, the orientational order-disorder transition of lyotropic liquid crystals tuned by the density, the conductor-insulator transition of dirty metals tuned by the disorder strength, localization-delocalization transitions in random matrices, percolation thresholds, or the Hausdorff dimension of sample paths of $\mathrm{SLE}_\kappa$.

The name `quantum phase transition' has -- somewhat unfortunately -- been given to a transition happening at zero temperature within the ground state space of a family of Hamiltonians. The drosophila of the subject is the one-dimensional transverse field Ising Hamiltonian,
\begin{equation*}
H_N(\lambda) = -\sum_{i=1}^{N-1} \sigma_i^x\sigma_{i+1}^x - \lambda\sum_{i=1}^N\sigma_i^z,\qquad \lambda\geq 0,
\end{equation*}
defined here on a chain of length $N$, namely on $\caH_N = \otimes_{i=1}^N \bbC^2$ with open boundary conditions. The matrices $\sigma_i^{x},\sigma_i^z$ are Pauli matrices acting on the $i$th site, i.e., tensored to the identity everywhere else. Minimizing the first term alone yields two possible product states of either eigenstates of $\sigma^x$, whereas the minimizing state of the second term has all spins in the $+1$ eigenstate of $\sigma^{z}$. Between these two extreme cases, $\lambda$ parametrizes a phase transition, which can be studied explicitly~\cite{Pfeuty}. On both sides of the critical parameter $\lambda_c$, the Hamiltonian has a spectral gap above the ground state energy which is uniformly bounded in the length of the chain, and correlations decay exponentially. However, as indicated by the limiting cases, the ground state is non-degenerate for $\lambda>\lambda_c$, while there is a double degeneracy whenever $\lambda<\lambda_c$. The spectral gap closes at $\lambda=\lambda_c$. One of the fascinating aspects of this model is the theoretical prediction of an emergent $E_8$ symmetry of the excitation spectrum at the critical point~\cite{ZamE8}, traces of which have recently been observed~\cite{Coldea} (for a thorough mathematical discussion of the matter, see the recent article~\cite{AMS}).

With this textbook example of a quantum phase transition in mind, let us consider the first, fundamental question: What is in fact a ground state phase?  For the purpose of this note, the attention will be restricted to systems with gapped ground states. In the Ising case -- as in other explicitly solvable models such as the $XY$-chain~\cite{LSM} -- the transition occurs when the gap closes. Reciprocally, there cannot be a phase transition without closing the spectral gap above the ground state. Hence the following definition~\cite{ChenGuWen, Hastings_Houches}: Two gapped Hamiltonians $H_0$ and $H_1$ are in the same phase if there exists a smooth path $H(s)$ for $s\in[0,1]$ such that $H_0=H(0)$, $H_1=H(1)$ and the spectral gap remains open, uniformly in the size of the system and in the parameter $s$. From a physical point of view, this implies that the respective ground states could be reached from one another by adiabatically following the given path, e.g., by manipulating external fields.

This Ising type of transition is quite similar to the classic phase transitions -- and indeed it corresponds to the $\beta\to\infty$ limit of it. The introduction of the toric code model in~\cite{KitaevModel} added a new dimension to the discussion of ground state phases. Its nearest neighbour interaction between spins-$1/2$ can be defined on an arbitrary two-dimensional lattice, with deep consequences for the structure of the ground state space. Indeed, when defined on a surface of genus $g$, the ground state degeneracy is equal to $4^g$, a property called `topological order'. Moreover, any local observable -- an operator that acts non-trivially only on a finite, contractible subset of the lattice -- reduces to a multiple of the identity when restricted to the ground state space. In this respect, a second definition is natural~\cite{ChenGuWen}: Two gapped ground states $\Omega_0$ and $\Omega_1$ are in the same phase if there exists a local unitary map such that $\Omega_1 = U\Omega_0$. 

Understanding the relation between these two definitions allows for a deep clarification of the notion of a gapped ground state phase which takes topological order into account.


We consider a countable number of quantum systems labelled by $x\in\Gamma$, where $\Gamma$ is equipped with a distance $d(\cdot,\cdot)$. Each point carries a finite dimensional Hilbert space $\caH_x$. The Hilbert space of a finite subsystem $\Lambda\subset\Gamma$ is the tensor product $\caH_\Lambda = \otimes_{x\in\Lambda}\caH_x$. The interaction between these `spins' is local in the sense that the Hamiltonian is of the form
\begin{equation*}
H_\Lambda(\lambda) = \sum_{X\subseteq\Lambda}\Phi(X,\lambda)\,,
\end{equation*}
where the self-adjoint operator $\Phi(X,\lambda)$ acts on $X$ only and $\Phi(\cdot,\lambda)$ decays in the size of the set $X$. For example, a finite range interaction would have $\Phi(X,\lambda) = 0$ if $\mathrm{diam}(X)\geq R$. We consider a family of local Hamiltonians parametrized by $\lambda$, e.g., the strength of an external magnetic field. We assume that $H_\Lambda(\lambda)\geq 0$. 

Such local Hamiltonians generate a dynamics $\tau^\Lambda_t(\cdot)$ on the algebra of observables characterized by a finite `speed of sound' as made precise by Lieb-Robinson (LR) bounds, see~\cite{LR} and the references therein in a previous edition of this Bulletin. Let $A$ and $B$ be two bounded operators acting on disjoint subsets $X$ and $Y$ respectively. Under mild assumptions on the structure of $\Gamma$ and with sufficient decay of the interaction, the following bound holds:
\begin{equation*}
\left\Vert [A, \tau_t^\Lambda(B)] \right\Vert\leq C(A,B) \exp\left[-\mu (d(X,Y) - v_\Phi|t|) \right],
\end{equation*}
where $v_\Phi$ is the Lieb-Robinson velocity, and $\mu>0$. 

The locality of the dynamics exhibited by the LR bound plays a central role in the following theorem, where we come back to the question of gapped ground states. For a finite $\Lambda$, let $\caS_\Lambda(\lambda)$ be the ground state space of $H_\Lambda(\lambda)$, i.e., the set of states such that $\rho(H_\Lambda(\lambda)) = 0$. Let $(\Lambda_n)_{n\in\bbN}$ be an increasing and absorbing sequence of finite sets converging to $\Gamma$. The set $\caS_\Gamma(\lambda)$ of states in the thermodynamic limit is then obtained by standard compactness arguments from~$\caS_{\Lambda_n}(\lambda)$.
\begin{thm}
Suppose that the family of interactions $\Phi(X,\lambda)$ is uniformly bounded, of class $C^1$ for $\lambda\in[0,1]$, and has finite range. If the spectrum of $H_\Lambda(\lambda)$ is characterized by a uniform lower bound $\gamma>0$ above the ground state energy, then there exists a cocycle of quasi-local automorphisms $\alpha_{\lambda,\lambda_0}^\Gamma$ of the algebra of observables such that
\begin{equation*}
\caS_\Gamma(\lambda) = \alpha_{\lambda,\lambda_0}^\Gamma(\caS_\Gamma(\lambda_0)),\qquad \lambda,\lambda_0\in[0,1].
\end{equation*}
\end{thm}
Note that the finite range assumption can in fact be relaxed to a sufficient decay~\cite{BMNS}. The automorphisms $\alpha_{\lambda,\lambda_0}^\Gamma$ are quasi-local in the sense that a strictly local observable is mapped to an almost local one, namely to one which is supported on a finite number of spins, up to an exponentially small correction. We note three important facts: 
\begin{enumerate}\setlength{\itemsep}{0pt}
\item $\alpha_{\lambda,\lambda_0}^\Gamma$ acts in the thermodynamic limit, i.e., on the algebra of quasi-local observables;
\item $\alpha_{\lambda,\lambda_0}^\Gamma$ is invariant under local symmetries of the interaction: If $\pi$ is an automorphism such that $\pi(\Phi(X,\lambda))
=\Phi(X,\lambda)$, for all $X\subset \Gamma$ and $\lambda\in[0,1]$, then $\pi$ is also a symmetry of $\alpha_{\lambda,\lambda_0}^\Gamma$, i.e., $\alpha_{\lambda,\lambda_0}^\Gamma\circ\pi =\alpha_{\lambda,\lambda_0}^\Gamma$;
\item $\alpha_{\lambda,\lambda_0}^\Gamma$ maps the complete set of ground states of one model, at $\lambda_0$, to the set of ground states of another model, at $\lambda$.
\end{enumerate}
In other words, the automorphism preserves the general structure of the ground state space. 

The spectral flow $\alpha_{\lambda,\lambda_0}^\Gamma$ is a generalization of the `quasi-adiabatic continuation' of~\cite{HastingsQuasi}, and all its properties follow from its explicit form. In fact, the theorem is constructive and the automorphism is obtained as the thermodynamic limit of a unitary conjugation defined on the algebra of local observables,
\begin{equation}\label{ThermoLim}
\alpha_{\lambda,\lambda_0}^\Gamma (A) := \lim_{n\to\infty} V^*_n(\lambda) A V_n(\lambda),
\end{equation}
where $V_n(\lambda_0)=1$ and $V_n(\lambda)$ solves a Schr\"odinger equation $V_n^\prime(\lambda)=\mathrm{i} D_n(\lambda)V_n(\lambda)$. The generator $D_n(\lambda)$ is given explicitly by
\begin{equation}
\label{D}
D_n(\lambda) := \int_{-\infty}^\infty dt\, w_\gamma (t) \int_0^t du\, \tau^{\Lambda_n}_u(H_{\Lambda_n}^\prime(\lambda))
= \sum_{Z\subset \Lambda_n}\Psi_n(Z,\lambda),
\end{equation}
where $w_\gamma\in L^1(\bbR)$ decays almost exponentially and has a Fourier transform which is supported in $[-\gamma,\gamma]$. The first equality is a definition. The second however requires the use of the LR bound. Indeed, it is the locality of~$\tau_t$ that ensures that $D_n(\lambda)$ can be cast as a local interaction $\Psi_n(Z,\lambda)$ with fast decay. This local structure implies in turn a LR estimate for the unitary dynamics~(\ref{ThermoLim}) in `time' $\lambda$. With this bound, it is a standard application to prove the existence of the thermodynamic limit $\alpha_{\lambda,\lambda_0}^\Gamma$.

On a side note, it is worth mentioning that the spectral flow is not restricted to the mapping of ground state spaces. In fact, for any family of operators with bounded derivative and a uniformly isolated spectral patch with associated spectral projection $P(\lambda)$, we have $P^\prime(\lambda) = \mathrm{i} [D(\lambda), P(\lambda)]$, where $D(\lambda)$ is defined by the integral in~(\ref{D}).

Now, what is the significance of this theorem for the question of gapped ground state phases of quantum spin systems? First of all, it clarifies and relates the two definitions of a phase, namely the one involving a gapped path of Hamiltonians and the one using a local unitary map: Paraphrasing, the ground state spaces of a smooth family of gapped Hamiltonians are related by a quasi-local automorphism of the algebra of observables. Furthermore, the locality of the map is made explicit and allows for its extension to infinite systems, a non-trivial but crucial step for the study of phase transitions. Also, including symmetries and symmetry breaking in the discussion is immediate.

Last but not least, the notion of local automorphic equivalence offers a natural way to describe topological phases. Topological order may have received various definitions, but the common feature is -- paradoxically -- that topologically ordered systems are locally disordered: No local order parameter can distinguish between the various ground states.
Topological order is revealed by considering the same interaction on different lattices. Hence the following proposition. First fix a set of lattices $\caL$. In one dimension, that would be $\bbZ$ and the two possible half-infinite chains with one boundary; in two dimensions, e.g., $\bbZ^2$, various half-planes, and closed surface of different genera. Consider two interactions $\Phi_0$ and $\Phi_1$ with corresponding Hamiltonians $H_{0,\Gamma}$ and $H_{1,\Gamma}$ that are gapped for all $\Gamma\in\caL$, with ground state spaces $\caS_{0,\Gamma}$ and $\caS_{1,\Gamma}$. Then, the two models are in the same ground state phase if for all $\Gamma\in\caL$ there exists a quasi-local automorphism $\alpha^\Gamma$ mapping $\caS_{0,\Gamma}$ to $\caS_{1,\Gamma}$. Hence, not only does the structure of the two ground state spaces in the bulk have to be similar -- for example of equal dimension -- but they must also depend on the underlying geometry and topology in the same way.

Let us briefly discuss the one-dimensional case. In~\cite{ChenGuWen, Schuch} it is concluded that all gapped, translation invariant, one-dimensional quantum spin systems without symmetry breaking belong to the same phase, and that they are in particular equivalent to a product state: There is no trace of topological order in one dimension. This may be true for the bulk phase, but fails to take into account the behaviour at the boundary. In~\cite{PVBS} we introduced a family of models that illustrate the role of edge states. These Hamiltonians all have a unique gapped ground state in the limit of an infinite chain, which is a product state indeed. However, when they are defined on the right half-infinite chain with a left boundary, there are $2^{n_L}$ ground states which can be interpreted as a bulk product vacuum upon which $n_L$ distinguishable particles can be added. At most one of each can bind to the edge without raising the energy, but a second one, in the bulk, represents an excitation above the spectral gap. A similar situation happens on the half-infinite chain with a right boundary, and ${n_R}$ particles. The exact `masses' of these particles are irrelevant: two such models are in the same gapped phase if and only if they have the same numbers $n_L$ and $n_R$. They are very simple representatives of the equivalence classes defining ground state phases and in fact, it is possible to prove that the famous AKLT model~\cite{AKLT} belongs to the class indexed by $n_L = n_R = 1$.

Before concluding, it should be noted that this leaves the question of the actual nature of topological order unanswered. According to the physics literature, it is a manifestation of long range entanglement as detected by the scaling behaviour of the entanglement entropy,
\begin{equation*}
S(\rho_X):=-\mathrm{Tr}\left(\rho_X\ln\rho_X\right),
\end{equation*}
where $\rho_X$ is the density matrix describing the restriction of the bulk state to the finite set $X\subset\Gamma$. The area law conjecture, proved only in one dimension~\cite{Area}, states that the entropy scales as the area of~$X$, $S(\rho_X)\leq C\vert \partial X\vert$, for general gapped quantum spin systems with finite-range interactions. Topological order is expected to manifest itself through universal subleading terms to that law, e.g., a constant in two dimensions, see~\cite{EntEnt} and references therein. A further promising generalization is the concept of entanglement spectrum~\cite{EntSpec}.

In this short note I have concentrated on the very definition of (gapped) ground state phases, and on the closely related problem of classifying them. Far reaching proposals including symmetries already exist in the literature, e.g.~\cite{CheGuWen2} for one-dimensional systems. The fundamental relation of topological order with non-local entanglement requires a rigorous analysis. The mathematical study of quantum phases and of the transitions between them is still at embryonic stages, and many physically relevant problems remain open.

\paragraph{Acknowledgements.} Work supported by the NSF Grant \#DMS-0757581.


\bibliography{Ref_Bulletin}
\bibliographystyle{unsrt}

\end{document}